# Sorting of particles suspended in whole blood

Stefan H. Holm[1]‡, Zunmin Zhang[2]‡, Jason P. Beech[1], Gerhard Gompper[2], Dmitry A. Fedosov[2*] and Jonas O. Tegenfeldt[1*]

1. Department of Physics and NanoLund at Lund University, Lund, SWEDEN
E-mail: jonas.tegenfeldt@ftf.lth.se

2. Theoretical Soft Matter and Biophysics, Institute of Complex Systems and Institute for Advanced Simulation, Forschungszentrum Jülich, 52425 Jülich, GERMANY

‡ These authors contributed equally to this work

[*] Corresponding authors

d.fedosov@fz-juelich.de

jonas.tegenfeldt@ftf.lth.se

## Abstract

*An important step in diagnostics is the isolation of specific cells and microorganisms of interest from blood. Since such bioparticles are often present at very low concentrations, throughput needs to be as high as possible. In addition, to ensure simplicity, a minimum of sample preparation is important. Therefore, sorting schemes that function for whole blood are highly desirable. Deterministic lateral displacement (DLD) has proven to be very precise and versatile in terms of a wide range of sorting parameters. To better understand how DLD performs for blood as the hematocrit increases, we have performed measurements and simulations for spherical particles in the micrometer range moving through DLD arrays for different flow velocities and hematocrits ranging from pure buffer to whole blood. We find that the separation function of the DLD array is sustained, even though blood cells introduce a shift in the trajectories and a significant dispersion for particles that are close to the critical size in the device. Simulations qualitatively replicate our experimental observations and help us identify fundamental mechanisms for the effect of hematocrit on the performance of the DLD device.*

## Introduction

Blood separation is an important procedure in biomedicine. As the blood sample drawn from a patient usually contains a plethora of biological markers, ranging from electrolytes, signaling molecules, proteins and nucleic acids to viruses, cells and microorganisms, an efficient sorting must be accomplished to minimize confounding effects of irrelevant cells or to enrich the bioparticles of interest for early and accurate blood diagnosis. This is typically done using centrifugation, which is a simple and well-established technique. However, it requires large sample volumes, making it difficult to integrate with lab-on-a-chip devices.

To significantly decrease the amount of sample needed for detailed analysis, several microfluidic approaches have been developed[1]. For example, acoustophoresis[2] and inertial focusing[3] offer high throughput but require diluted samples. Particle margination or migration in blood flow is an interesting approach that mimics white blood cell (WBC) migration in small blood capillaries[4] and works best at red blood cell (RBC) hematocrits (volume fraction of RBCs) similar to those in whole blood. The margination effect has been used to separate out stiffer infected blood cells for malaria detection[5]. However, it is quite limited in scope.

Deterministic lateral displacement (DLD), on the other hand, has been shown to be a highly versatile mechanism for particle sorting in microfluidics[6]. The underlying mechanism, which distinguishes between two different types of motion for smaller and larger spherical rigid particles in relation to a critical size, is well understood[7]. While the first studies using DLD have demonstrated highly precise sorting of microspheres as a function of size, subsequent developments have demonstrated possibilities for sorting with respect to other relevant particle properties, *e.g.* shape[8-11], deformability[10], viscosity contrast[12] and dielectric properties[13]. Numerous research efforts have further advanced the understanding of the effects of device geometry [6, 7, 9-11, 14-26] and particle properties [8-13, 16, 17, 20, 21, 27-33] on sorting. For instance, Davis and Huang [6, 27] have shown that large flow velocities tend to improve the resolution of separation for solid particles because fast flows decrease the influence of diffusion. However, as first discussed by Davis *et al.* [27] and later demonstrated by Beech *et al.* [10], sufficiently high shear forces at high flow rates may significantly deform soft particles, changing their effective size.

The first separations performed on blood [27] showed the relevance of DLD for diagnostics. Subsequent studies have demonstrated sorting of many biologically and diagnostically relevant components[8-11, 21, 25, 29, 33-46]. A question that remains largely unexplored is the effect of blood hematocrit, on the performance of these devices for sorting and detection of particles suspended in blood. While whole blood sorting has been demonstrated[47], the concern remains that non-deterministic interactions between particles at high concentration affect their trajectories, having deleterious effects on the relevant threshold sizes and resolution of the sorting. Feng *et al.* developed modifications to devices that allowed for increased particle concentrations during enrichment [48]. Vernekar and Krüger performed 3D simulations of RBCs that showed that the trajectories of RBCs in DLD devices are affected by their hematocrit [49]. To optimize sorting of particles from whole blood, it is necessary to understand how the trajectories of particles moving through a DLD array are affected not only by the device geometry but by interactions with the RBCs with which they are travelling and consequently by the hematocrit, or the volume fraction of RBCs in the sample.

We perform experiments and simulations to elucidate the effect of hematocrit on the sorting efficiency of DLD devices. Careful experiments are performed using a single-section device (with a constant row shift) to avoid overlay of different displacement mechanisms. Simulations, which explicitly represent deformable RBCs embedded in a Newtonian fluid, are employed to elucidate the mechanisms induced by finite hematocrit. We use spherical particles with diameters in the range 2 μm to 10 μm as models for the plethora of biological markers mentioned above, to facilitate a systematic study of the effect of particle size. Our main result is that sorting remains possible up to surprisingly high hematocrit values. However, the sorting is no longer completely deterministic because the presence of RBCs leads to increased dispersion. The dispersion by RBCs has the strongest effect on particles that are close to the critical size of the DLD array in the device. This study provides an understanding of the mechanisms, which govern the dispersion effect of hematocrit on particle sorting and quantifies the performance of DLD devices for particle isolation from whole blood and diluted blood.

## Materials and Methods

*Microfluidic devices*

Two devices were used in this work. Device 1 is a chirped array with critical diameters raging from, $D_C = 2.86$ μm to $D_C = 9.04$ μm. Device 2 is a uniform array with one critical size, $D_C = 6.5$ μm. Design details are described in the ESI.

Devices were fabricated from PDMS (Sylgard 184, Dow Corning, Midland, MI, USA) using standard replica molding[55] of SU8 (SU8-2010, Microchem Corp., MA, USA) structures on a 3” silicon substrate. Designs were drawn in L-Edit 11.02 (Tanner Research, Monrovia, CA USA), the photomask printed by Delta Mask (Delta Mask, Enschede, The Netherlands), and the master fabricated using UV-lithography (Karl Suss MJB4, Munich, Germany). To facilitate demoulding from the master it was coated with an anti-sticking layer, 1H,1H,2H,2H-per-fluorooctyl-trichlorosilane (ABCR GmbH & Co. KG, Karlsruhe, Germany) [56]. PDMS casts were oxygen-plasma bonded to microscope slides spun with a thin layer of PDMS (Plasma Preen II-862, Plasmatic Systems Inc., North Brunswick, NJ, USA) to obtain devices with all internal surfaces consisting of PDMS. Immediately after bonding, the devices were filled with solution of a 0.2 % (w/v) PLL(20)-g[3.5]-PEG(2) (SuSoS AG, Dubendorf, Switzerland) in deionized water and left for 20 minutes before subsequence flushing in order to minimize the adhesion of cells to the walls. Silicon tubing with 3mm inner and 5mm outer diameter (228-0725 and 228-0707, VWR International LLC, Radnor, PA, USA) were glued (Elastosil A07, Wacker Chemie AG, Munich, Germany) onto the device as reservoirs.

*Measurement setup*

An MFCS-4C pressure controller (Fluigent, Paris, France) was used to control the overpressure at the inlets and the outlet reservoirs were kept at ambient pressure. Images were captured through an inverted Nikon Eclipse Ti microscope (Nikon Corporation, Tokyo, Japan) with an Andor NEO sCMOS camera (Andor Technology, Belfast, Northern Ireland) and Lumencor SOLA light engine$^{TM}$ (Lumencor Inc, OR, USA) with FITC, TRITC and DAPI filters, or brightfield and 4x (Nikon Plan Fluor NA0.13), 10x (Nikon Plan Fluor NA0.3) and 60x (Nikon Fluor water immersion NA 1.0) objectives were used and all movies acquired at 10 frames per second.

The results for both experiments and simulations are characterized by the displacement index, which is defined as the ratio of the lateral displacement of particles per post to the row shift. For the ideal displacement mode, it should be one as the particles are displaced laterally from their original stream to flow along the gradient of post array. For the ideal zigzag mode, it will be close to zero as the particles move in the direction of flow without net lateral displacement.

*Reagents and blood samples*

Experiments were performed by flowing suspensions of polystyrene microspheres of diameters 1.8±0.05 µm and 4.72±0.15 µm (#17687, #18340, Polysciences Inc.), 6.53±0.62 µm and 9.80±0.67 µm (#35-2, #36-3, Duke Scientific Corp.) and 7.81±0.11 µm (PS-FluoGreen-Fi220, microParticles GmbH), while simultaneously flowing solutions of human blood into the background inlets at hematocrits of 0, 5, 15, 30 and 45% (whole blood). The lateral positions (degree of displacement) of the microspheres were measured at the end of the separation array and a displacement index (0 = zigzag mode and 1 = displacement mode) was calculated.

An automated script was used to analyze the exit distribution of the sorted particles. The script is based on an analysis of intensity of the fluorescent particles after compensation for background. Each data point is derived from more than 100 particles. Due to the larger concentration in terms of number or particles per unit volume, the smaller particles were sampled to a greater extent than the larger particles.

*Simulations*

Simulations took place with design parameters corresponding to device 2 (see ESI). To model fluid flow within a microfluidic device, we employ the dissipative particle dynamics (DPD) [57, 58] which is a mesoscopic hydrodynamics simulation technique. In this method, the simulated system consists of a collection of $N$ particles with mass $m_i$, position $r_i$ and velocity $v_i$, and each individual particle represents a cluster of atoms or molecules. The dynamics of the DPD particles is governed by Newton's second law of motion, and the total force acting between two particles within a selected cut-off ($r_C$) region is a sum of pairwise conservative, dissipative and random forces.

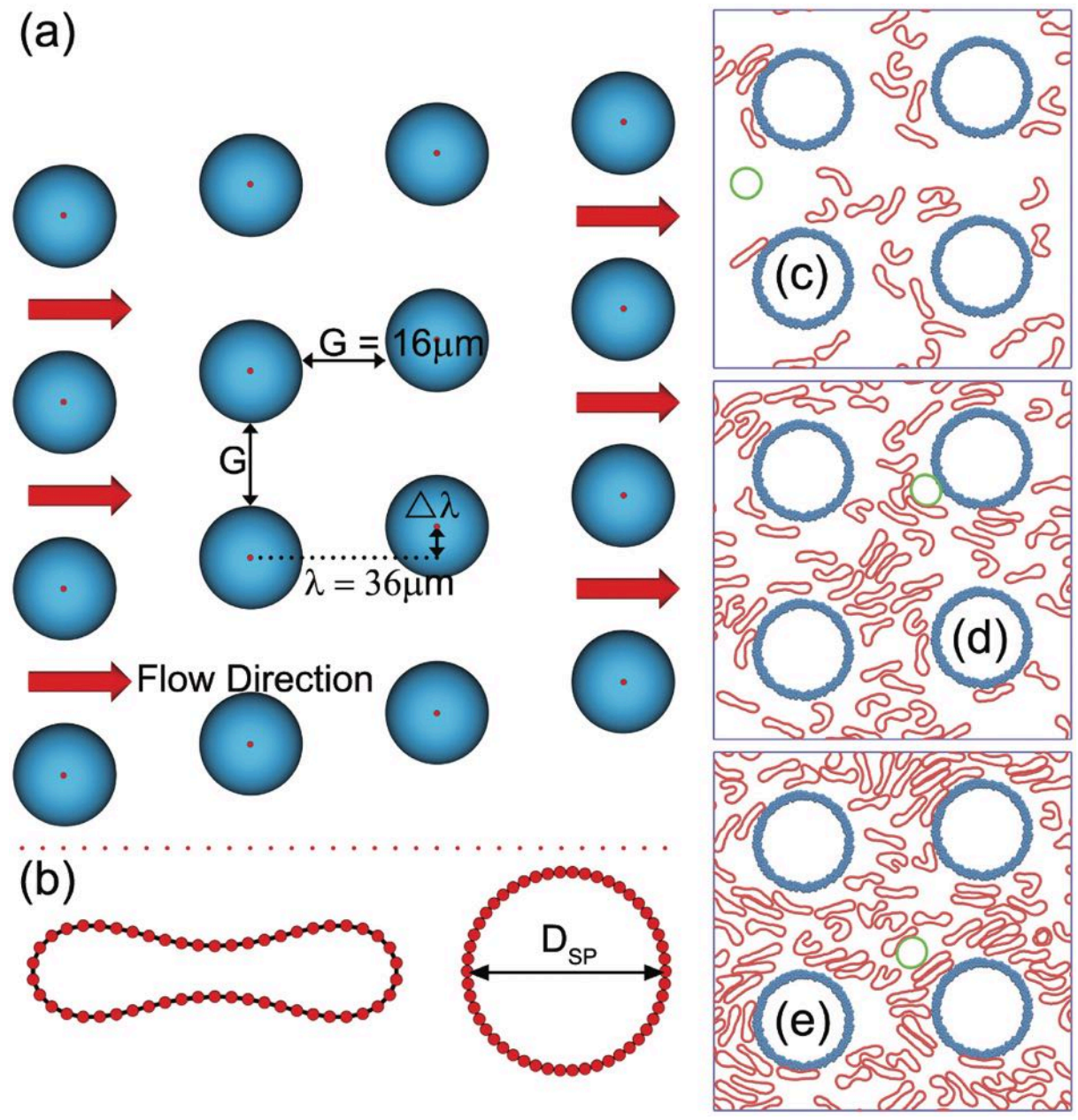

*Figure 1. Schematic illustrations of the simulated DLD system: (a) DLD array with circular posts. The geometry is defined by the post center-to-center distance λ, the post gap G, and the row shift Δλ. (b) 2D models of RBCs and rigid spherical particles. (c-e) Instantaneous snapshots of typical fluid structures at different hematocrits: (c) 11.31%, (d) 22.62%, and (e) 33.93%. DLD device parameters are λ = 36µm, Δλ = 2.4µm, and G = 16µm corresponding to device 2 (see ESI).*

Modelling and simulation of blood flow on the cellular level have made enormous progress in recent years[59]. We employ here two−dimensional (2D) simulations to study the effect of the hematocrit on the separation of rigid spherical particles in DLD devices with circular post arrays. Both RBCs and rigid spherical particles are modelled as closed bead-spring chains with $N_V$ particles ($N_V = 50$ for RBCs and $N_V = 30$ to 60 for rigid spherical particles depending on their size) connected by $N_S = N_V$ springs, as illustrated in **Error! Reference source not found.**. The bonding potential, bending energy, and area constraint are applied to control the shape and rigidity of the cells and particles[20]. These models have been validated to be able to capture the essential physical features required to correctly describe the motion of particles in conventional DLD devices with circular posts [20].

The simulated domain, which contains four posts, is periodic in both directions[20]. However, boundary conditions in the x-direction (the flow direction) are subject to a constant shift Δλ in the y-direction in order to mimic the shift between two consecutive rows of posts. Therefore, a shift is introduced for every boundary-crossing event. The flow in the x-direction is driven by a force applied to each fluid particle in order to control the flow rate. To enforce no net flow in y-direction, which occurs in the device due its side walls, a force in the y-direction is also applied to each fluid particle. This force is adapted separately for every hematocrit to satisfy the no-net-flow condition.

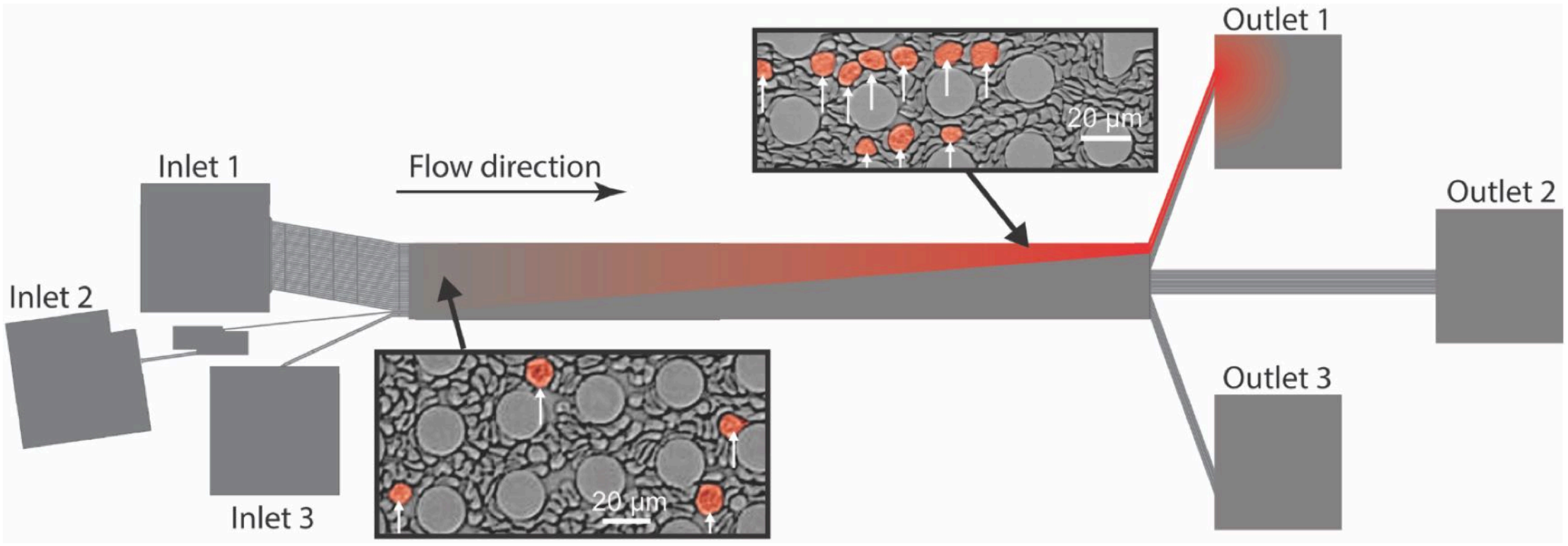

*Figure 2. Enrichment of WBCs in whole blood (45% hematocrit) using a DLD device (device 1, see ESI) with 13 separate sections, corresponding to $D_C$ ranging from approximately 3 µm to 9 µm in steps of 0.5 µm[11]. Left image shows an elevated concentration of WBCs in the middle of the device and the right image shows an even larger concentration at the end. WBCs are indicated by white arrows. Corresponding videos are available in the ESI.*

## Results and Discussion

### *Proof of principle: WBC enrichment in whole blood using a standard DLD device*

To demonstrate the usefulness of DLD sorting at high hematocrit, Figure 2 shows that WBCs can readily be enriched from a whole blood sample using a standard DLD device (device 1, see ESI). The DLD device consists of 13 separate sections with $D_C$ ranging from approximately 3µm to 9µm. Full details of the device design can be found in the electronic supporting information (ESI 1) and in Ref.[11]. A rough estimate based on cell counting at the end of the device, gives us an approximate ratio of one WBC per five RBCs. This means that an enrichment factor of about 200 is achieved in comparison to the physiological ratio of one WBC per 1000 RBCs in whole blood.

We used the same device (device 1, see ESI) to study the displacement of polystyrene microspheres. Figure 3 presents outlet distributions of the particles for different hematocrits, indicating that the effective size of the particle decreases as the hematocrit is increased, *i.e.* lateral displacement of particles decreases. At the same time, the width of the distribution increases with increasing hematocrit. These results suggest that sorting must still be possible at high hematocrits, even though RBCs at finite hematocrit lead to significant changes to particle trajectories when compared to those at zero hematocrit.

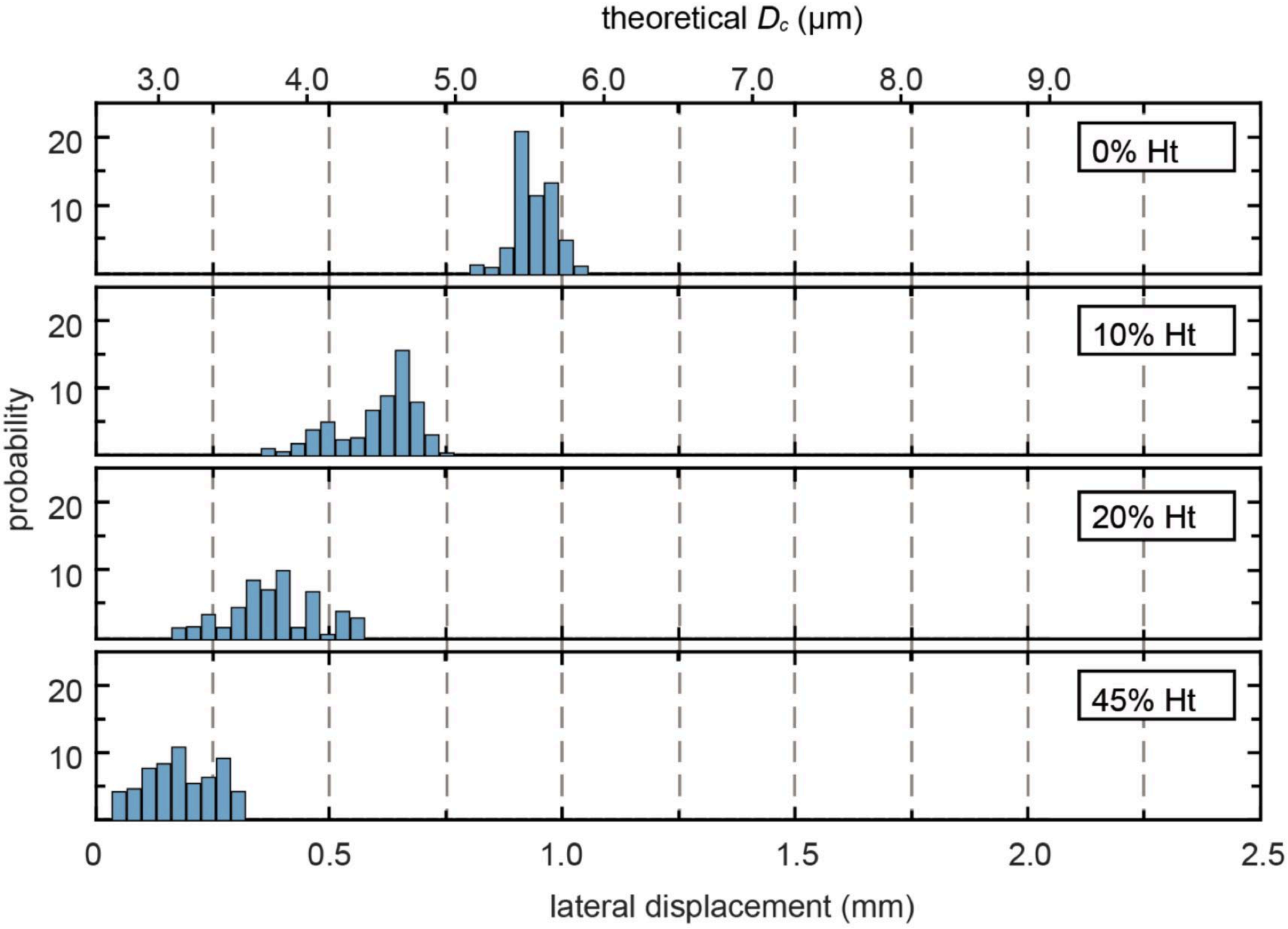

*Figure 3. Experimental outlet distributions of polystyrene particles (diameter 4.87±0.28µm) using the DLD device illustrated in Figure 1 (device 1, see ESI) for various hematocrits. The higher the hematocrit, the smaller the effective size of the polystyrene beads. On the top the corresponding critical sizes are given for the different outlet positions. This provides a scale for the effective sizes of the sorted particles.*

*Particle displacement in a single-section DLD with constant row shift*

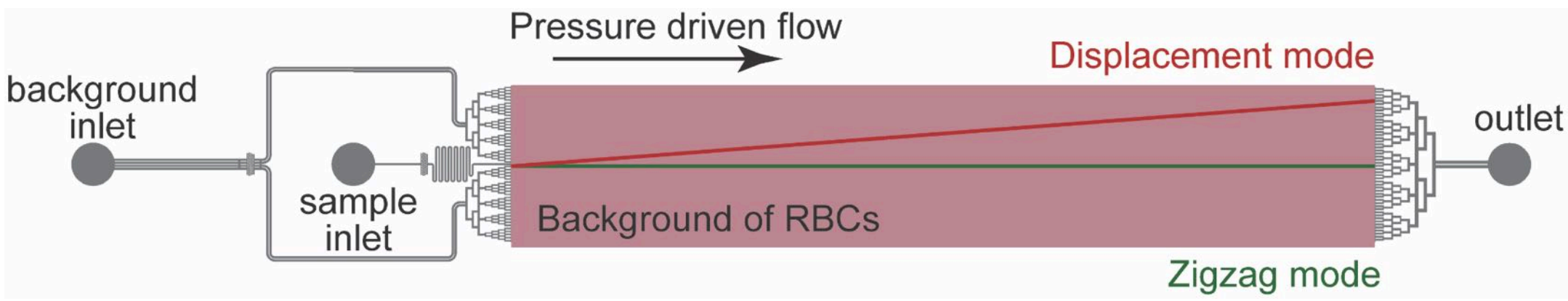

*Figure 4. DLD device with a single section having a constant row shift (device 2, see ESI). Post diameter 20 µm, height 15µm, gap size G = 16 µm, periodicity N = 15, corresponding to a row periodicity length $\lambda = 36\,\mu m$ and a row shift $\Delta\lambda = 2.4\,\mu m$. The device has two inlets: one central inlet for the microspheres and one inlet for the blood. The latter is guided to two wide flow streams on either side of the particle flow stream.*

In order to better understand the effect of RBCs on the trajectories of suspended particles, a new device has been built with a carefully selected row shift $\Delta\lambda = 2.4$ µm in a single section to avoid the overlay of different motions in integrated multi-sections, Figure 4 (device 2, see ESI). With a sufficiently large critical size, the RBCs inherently move in the zigzag mode and therefore exhibit a homogeneous cell distribution throughout the device. A systematic study can now be performed to examine the sorting performance for a variety of particle sizes and hematocrits.

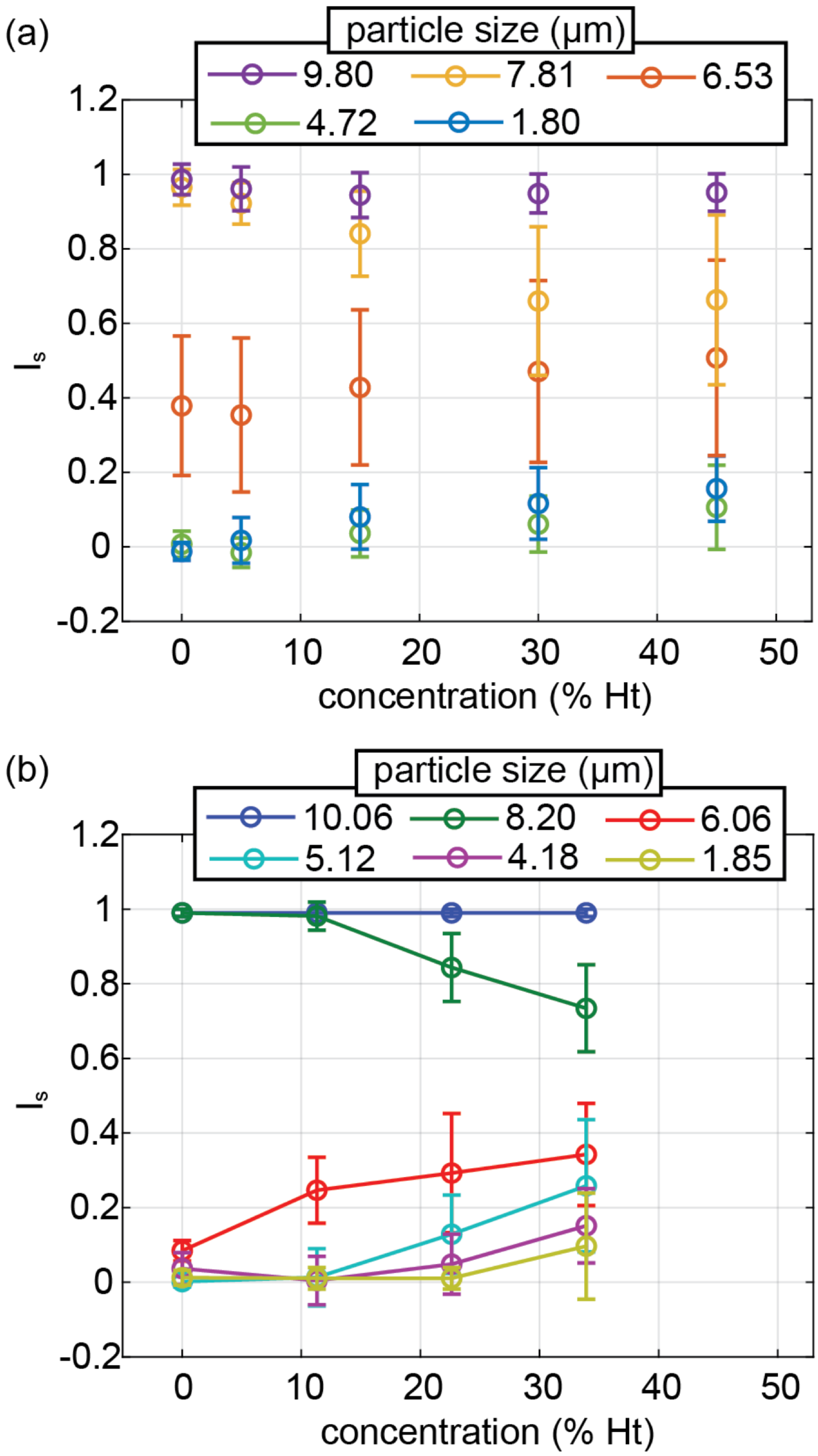

*Figure 5 Effect of hematocrit on the displacement of microspheres (device 2, see ESI). (a) The displacement index as a function of hematocrit obtained from the experiments at 100mBar driving pressure for five different microsphere diameters, $D_P$. (b) The displacement index from 2D simulations as a function of hematocrit ($H_t$) for six different microsphere diameters, $D_P$.*

As a reference, we have performed standard DLD separation of microspheres of various sizes (see ESI 3 for characterization of microsphere sizes) in an aqueous buffer solution without RBCs, *i.e.* 0% hematocrit. To quantitatively assess the separation efficiency of various particles, a displacement index, $I_S$, is defined as the average displacement of particles per post row divided by the row shift, which equals zero for a neutral zigzag mode (particles following the liquid flow) and unity for a perfect displacement mode (particles displaced by the device in relation to the liquid flow). Figure 5(a) shows that spheres with diameters 1.80 µm and 4.72 µm follow the average fluid flow direction unperturbed by the effects of the DLD array (zigzag mode). In contrast to the small particles, spheres with diameters 7.81 µm and 9.80 µm travel through the entire device in displacement mode. Spheres with diameter 6.53 µm have a mean displacement index of 0.4 indicating that the real critical size, $D_C$, is slightly larger than 6.53 µm, which in turn is larger than the nominal $D_C = 6.11$ µm. Interestingly, their trajectories have a much broader distribution than the other microspheres. This behavior corresponds to a so-called mixed mode, which has been observed both in experiments and simulations [6, 20, 23, 50]. In addition, two technical issues might be important here when the size of beads is close to the critical size of devices. Firstly, the size distribution of the beads spans the range below and above the critical size, and secondly the effect of any defects in the DLD array is amplified.

From the experiments where the microsphere particles are mixed in blood (Figure 5(a)), four different regimes can be identified (representative videos are available in the ESI):

*a)* For $D_P >> D_C$, no effect of the RBC background is observed, since particles with $D_P =$ 9.80 µm consistently exhibit the displacement mode characterized by the displacement index of 1.0.

b) When $D_P$ is close to and slightly above $D_C$, the displacement index decreases as the hematocrit is increased, which is illustrated by the data for $D_P$ = 7.81 µm in Figure 5(a).

*c)* When $D_P$ is close to and slightly below $D_C$, the increase of hematocrit results in an increase of the displacement index, as seen for $D_P =$ 6.53 µm. Note that with a displacement index < 0.5 for Ht = 0%, this particle size is considered less than the critical size.

*d)* For $D_P << D_C$, the displacement index of particles is nearly unaffected by the presence of RBCs. However, there is a slight increase in displacement index and the outlet distribution is wider with increasing hematocrit.

Clearly, the effect of hematocrit is strongest for particles whose size is close to the critical size of the device. It is intuitive to expect that suspended particles experience some degree of dispersion due to interactions or "collisions" with flowing RBCs. Higher hematocrits induce more collisions between the RBCs and particles, which increasingly disturbs the lateral displacement of the particles and in turn affects the separation efficiency. This proposition is also consistent with the observation in Figure 2 that the outlet distribution broadens with increasing hematocrit.

*Underlying mechanisms*

In order to elucidate the mechanisms of how hematocrit affects particle trajectories in DLDs, simulations of corresponding two-dimensional systems (see Materials and Methods) have been performed to provide a direct and qualitative understanding. Results for the displacement index as a function of hematocrit for various particle sizes are shown in Figure 5(b). The simulations are consistent with the trends observed in the experiments in Figure 5(a). For representative examples from the simulations, see videos in the ESI.

The following mechanisms are found to be at play.

1) *Shear-flow induced collisions and RBC pressure*: The local gradients in flow velocity result in frequent collisions among RBCs and between RBCs and particles, which lead to an effective pressure proportional to $H_t$, $\dot{\gamma}$, and $\eta_{eff}$, where $\dot{\gamma}$ is the local shear rate and $\eta_{eff}$ is the effective viscosity of the suspension[51]. In homogeneous channels, this pressure reduces the thickness of the RBC-free layer at the channel walls. For our DLD arrays, this pressure has several implications. It increases with increasing hematocrit, generates a more homogeneous distribution of particles in the available volume (Figure 6(a-c)), and pushes RBCs and particles closer to confining surfaces (Figure 7). Shear-induced collisions also affect particle margination or migration toward the walls[52].

2) *Layering of RBCs at post surfaces*: An immediate consequence of the shear-flow induced RBC pressure is the formation of an RBC-rich layer at confining surfaces. In addition, in DLD arrays the flow is not always tangential to the post surface, but in some locations has a strong perpendicular component. This enhances the layering effect in incident flow regions. The distributions of RBCs for various hematocrits are shown in Figure 7 and clearly demonstrate this effect.

3) *Go with RBCs*: An individual RBC at low hematocrit moves in zigzag mode through the device. Particles collide with the RBCs frequently and are thereby dragged into similar trajectories (Figure 8). With increasing hematocrit, RBCs also attain an increasing displacement component (Figure 8), which is related to the more homogeneous RBC distribution in the DLD.

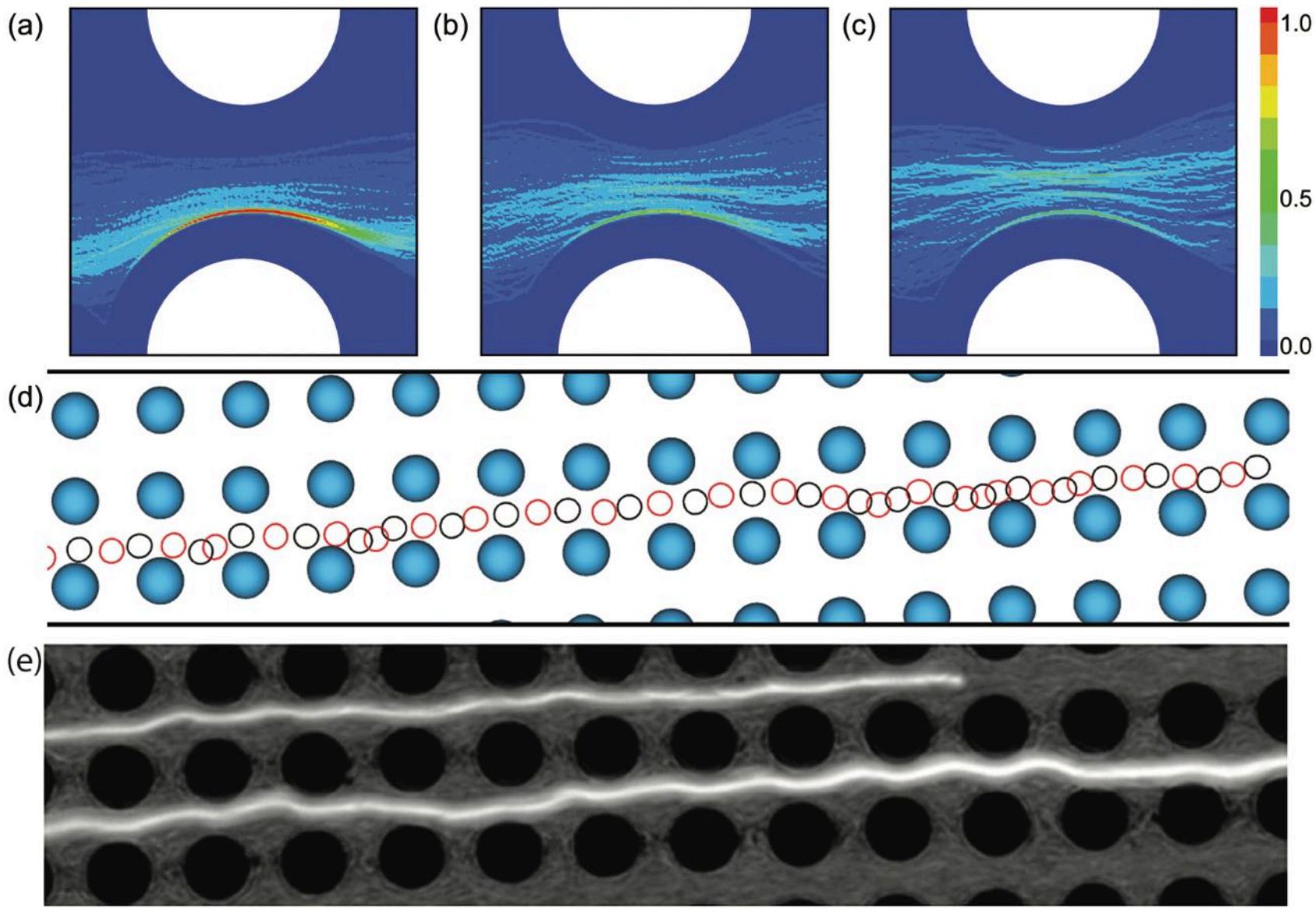

*Figure 6. Bumpy displacement trajectories (simulation and experimental, device 2, see ESI) of the big particles ($D_P$ = 10.06μm) in the presence of RBCs. (a-c) Simulated distributions of the suspended particles (center of mass) between two posts for Ht =11.31%, Ht = 22.62%, and Ht = 33.93%, respectively. (d) Simulated snapshot of a typical bumpy trajectory of the particle with $D_P$ = 10.06μm at Ht = 22.62%. (e) Experimental trajectories of fluorescent microspheres with a diameter $D_P$ = 9.8μm at 30% hematocrit.*

The consequences of these mechanisms on the four different particle-size regimes are discussed further below.

**a) $D_P >> D_C$.**

For large particles, both experiments and simulations in Figure 5 show that RBCs have no effect on the particle trajectories. Obviously, such large particles follow the displacement mode and any additional dispersion due to interactions with RBCs does not alter the displacement mode. Figure 6(d) and Figure 6(e) illustrate typical trajectories of large particles from simulations and experiments at moderate hematocrits. An interesting observation from these trajectories is that the particle often is displaced away from the post, as if it bumps off the post. This bumpy behavior is mediated by RBCs, since in an RBC-free suspension the displacement trajectory of the particle follows perfectly the post-array lane, where the particle 'touches' every subsequent obstacle. Furthermore, Figure 6(a-c) further reveals that the distribution of suspended particles gradually shifts away from the posts toward the gap center with increasing hematocrit, due to the formation of a high-density RBC layer at the post surface (Figure 7). This means that the displacement trajectory becomes increasingly bumpy with increasing hematocrit, wherein the particles irregularly move up and down within a lane between posts due to frequent collisions with the RBCs.

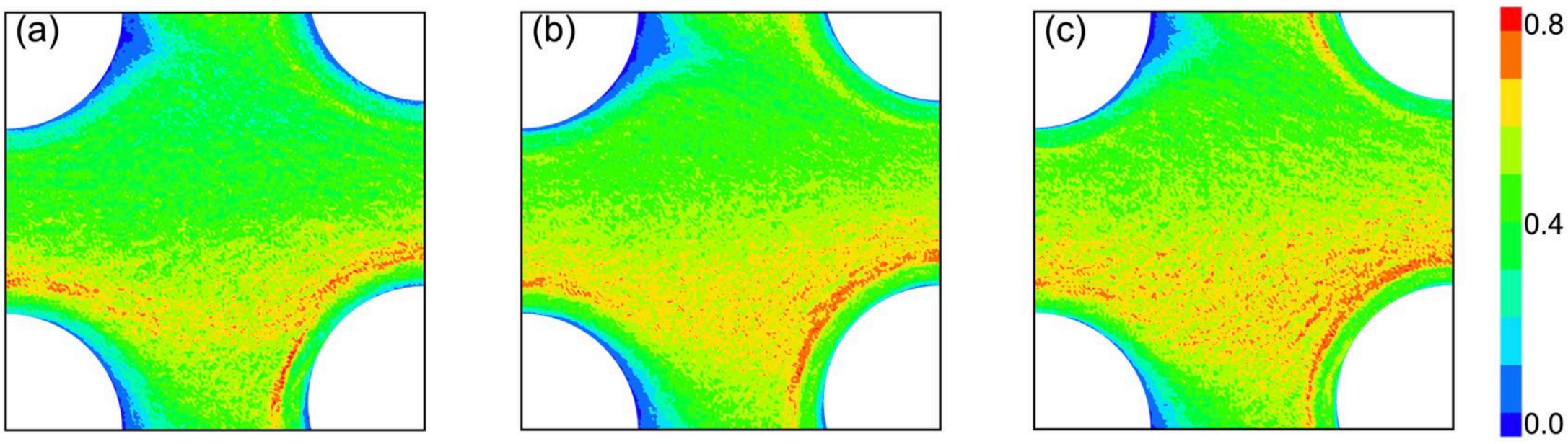

*Figure 7. Simulated distributions of the membrane particles of RBCs at hematocrits of (a) 11.31%, (b) 22.62% and (c) 33.93%. Membrane particles represent RBC surfaces and thus, their distribution is similar but not identical to local hematocrit. The distributions are averaged over several thousands of frames from long simulation trajectories and normalized by the maximum density.*

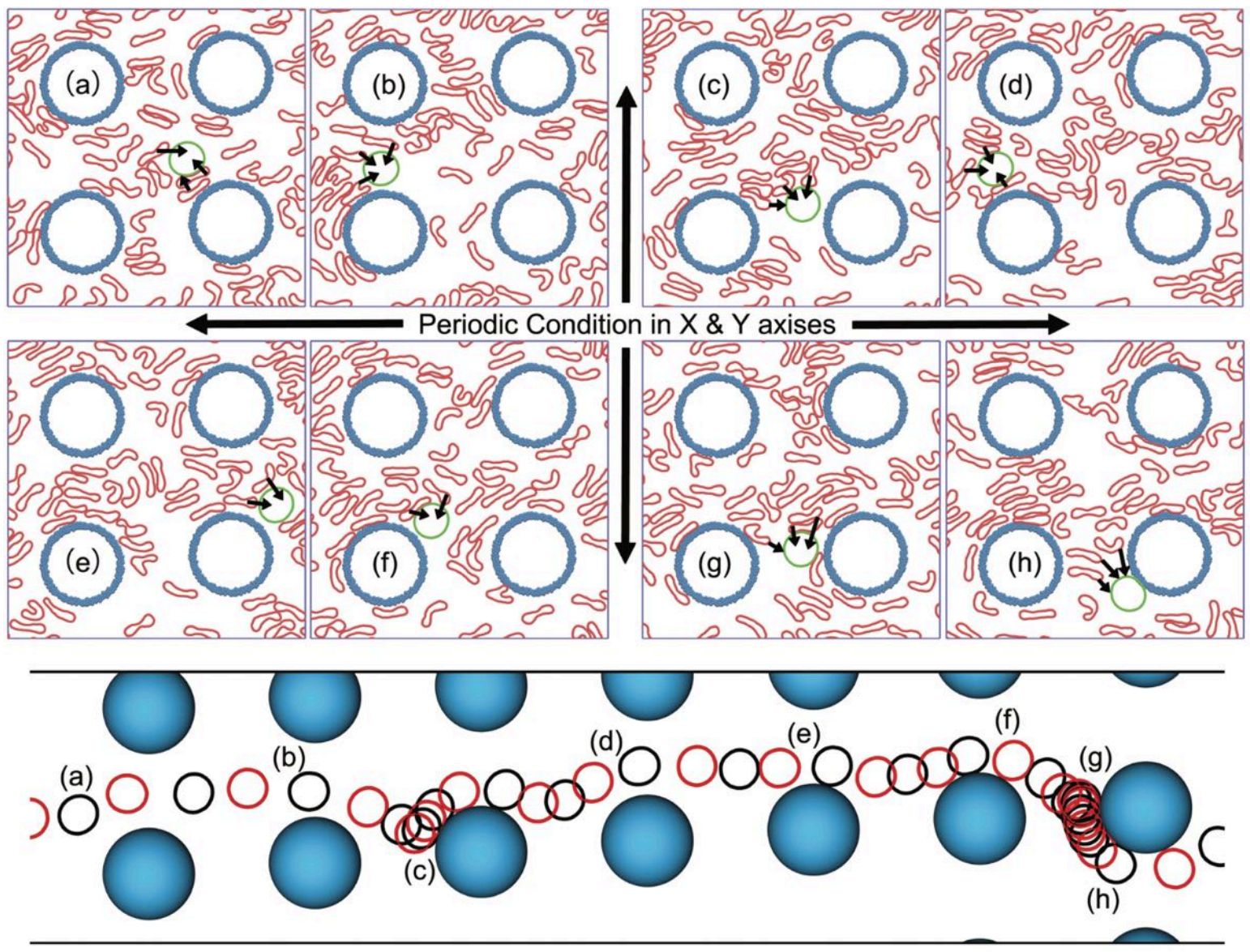

*Figure 8. Snapshots of typical flow behavior of RBCs and suspended particles (simulation), where a zigzag motion of the particle is promoted due to interactions with RBCs. Here, $D_P$ = 8.02μm > $D_C$ and Ht = 22.62%. Different snapshots are marked by letters a-h, corresponding to the particle positions in the trajectory displayed in the lower figure.*

**b) $D_P$ is close to $D_C$, with $D_P > D_C$.**

Particles are expected to exhibit a displacement mode in the buffer solution without RBCs for $D_P > D_C$. However, at high enough hematocrit for particles whose size is close to $D_C$ the displacement index is found to decrease to be significantly smaller than 1.0, as shown in Figure 5, implying the appearance of an induced zigzag event. Interactions with RBCs increase effective lateral diffusion of the particles, leading to an increased cross-streamline migration. Here, the collisions with RBCs are strong enough to occasionally promote lateral displacement of the particles into zigzagging. Such interactions are illustrated in Figure 8, where it is apparent that the particle encounters numerous collisions with RBCs. Snapshots in Figure 8(g-h) show how several RBCs drive and accompany the particle into the zigzag mode. The "go-with-RBCs" effect obviously depends on the hematocrit, since the zigzagging frequency increases with an increase of hematocrit, which is supported by the reduced displacement index for particles with $D_P \approx 8$ μm in Figure 5. However, it is clear that the interactions with RBCs favoring zigzag motion are not strong enough to force the largest particles into zigzagging and they remain in the displacement mode.

Note that the 'bumping' effect of RBCs, due to their layering near the posts illustrated in Figure 7, occurs here as well, which can be seen in Figure 8(a, d). Therefore, both 'bumping' and "go-with-RBCs" effects are present and their interplay influences particle trajectories. For particles whose size is close to $D_C$ and $D_P > D_C$, the "go-with-RBCs" effect is more prominent, resulting in a mixed mode with a displacement index smaller than unity.

**c) $D_P$ is close to $D_C$, *with* $D_P < D_C$.**

Particles naturally assume the zigzag mode without RBCs for $D_P < D_C$. With increasing hematocrit, the displacement index for such particles increases, as shown in Figure 5. Again, we need to consider the two effects of RBCs: (i) "layering of RBCs at post surfaces" that favors the displacement mode and (ii) "go-with-RBCs" that promotes zigzagging. Since the displacement index for particles with $D_P \approx 6$ µm increases, as the hematocrit is elevated, the frequency of zigzagging events must decrease. This indicates that the 'bumping' effect is dominant here, resulting in a decreased frequency of lane swapping events with increasing hematocrit.

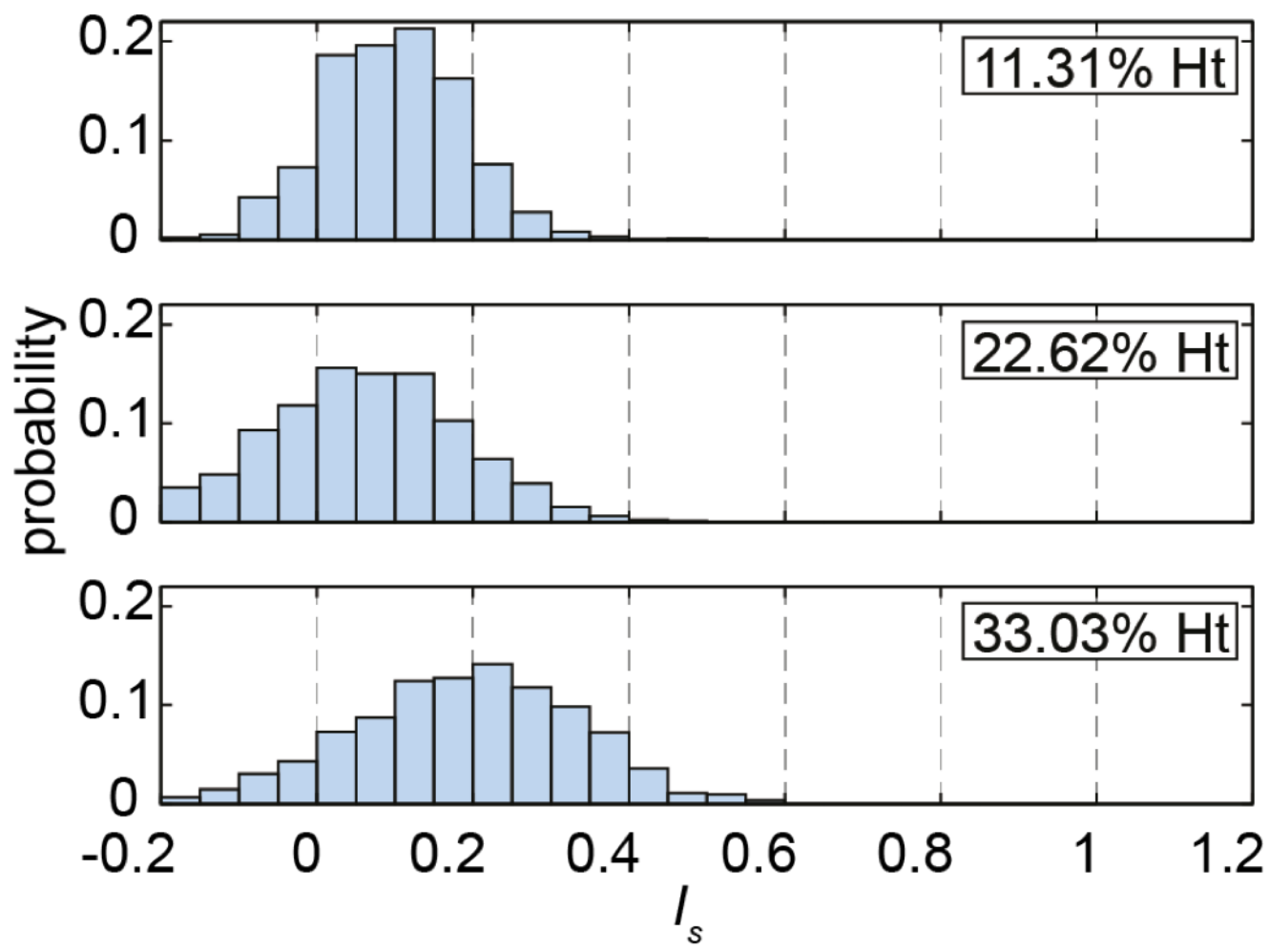

*Figure 9. Simulated outlet distributions of RBCs without any added particles as a function of the displacement index $I_S$ at different hematocrits. It shows how hematocrit affects the overall traversal of single RBCs through DLD device 2.*

**d) $D_P \ll D_C$.**

When the particle size is very small, its average trajectory remains nearly unaffected by RBCs and follows neutral zigzag mode. This indicates that the 'bumping' and 'go-with-RBCs' effects are approximately balanced. The displacement index for small particles in Figure 5(a) has a slight increasing tendency as the hematocrit is elevated, indicating that the 'bumping' effect may modestly decrease the frequency of zigzagging.

*Outflow distributions.*

Average outflow positions of the particles can be easily estimated from the displacement index in Figure 5, as they characterize the average particle displacement per one row of posts. To examine the effect of hematocrit on the width of outlet distributions, particle positions at the device exit were monitored. Figure 10 shows particle outlet distributions from simulations and experiments for various particle sizes. Simulated distributions for $D_P$ = 6.06 µm are qualitatively similar to experimental results for $D_P$ = 5 µm. In these cases, the outlet distributions are quite narrow in the absence of RBCs, corresponding to a nearly deterministic zigzag motion. With increasing hematocrit, both simulated ($D_P$ = 6.06 µm) and experimental ($D_P$ = 5 μm) distributions shift to a larger average particle displacement and significantly widen, since RBCs introduce dispersion to particle trajectories. Therefore, RBCs destroy the strictly deterministic nature of particle trajectories observed in a buffer solution. The dispersion effect of RBCs may also be non-monotonic, as the distribution for $H_t$ = 33.93% from simulations appears to be slightly narrower than for $H_t$ = 22.62%. Clearly, the width of outlet distributions depends on the length of a device or the total number of rows, since the dispersion has a cumulative effect. Thus, outlet distributions are expected to widen when the length of a DLD device is increased. Figure 10 also presents experimental particle outlet distributions for $D_P$ = 7 μm and $D_P$ = 8 μm. In case of $D_P$ = 8 μm with increasing hematocrit, the average particle displacement within the device slightly decreases, consistently with the displacement index in Figure 5(a) and the outlet distributions substantially widen. The outlet distributions for experiments with $D_P$ = 7 μm are very broad for all hematocrits (even without RBCs), because the particle diameter is very close to the critical size of the device. Thus, the presence of RBCs leads to significant widening of outlet distributions with the strongest effect for particles whose diameter is close to the DLD critical size.

It is also interesting to take a look at the trajectories of each single RBC. Figure 9 shows outlet distributions for RBCs at different hematocrit values. The width of RBC distributions increases with increasing hematocrit, consistently with the width of outlet distributions for suspended particles. Furthermore, there is a tendency for the RBCs to increase the displacement index as hematocrit is elevated, which can be seen from a shift of the distributions toward more positive values in Figure 9. The effective size of RBCs is smaller than $D_C$, so that the slight increase of the displacement index with increasing hematocrit is consistent with the analogous effect for small particles ($D_P \ll D_C$) in Figure 5. A non-zero displacement index for RBCs results in their slow positive migration within the DLD device. Thus, for a sufficiently long device, the distribution of RBCs far enough from the inlet may not be uniform with a varying local hematocrit across the device cross-section.

*Other potential contributions affecting particle trajectories*

The 'bumping' and 'go-with-RBCs' effects of RBCs discussed above provide an intuitive explanation for the behavior of suspended microspheres in DLD devices. However, this clearly does not exclude the possibility that other contributions affecting particle trajectories might be present. For instance, particle margination or migration toward walls of the device [53, 54] may contribute to the lateral motion of the particles. Due to the low lateral speed of particle margination, it is typically demonstrated in long channels. For example, Hou *et al.* used 5mm long channels [4]. For particles moving between posts, there is no obvious net build-up of lateral translation, making their margination not very probable. Furthermore, the particles are partially repelled from the posts due to an increased local hematocrit around the posts (*i.e.* the 'bumping' effect, Figure 7). Finally, interactions or collisions of microspheres with RBCs also represent one of the margination mechanisms, often referred to as shear-induced diffusion[52]. Therefore, it is not possible to make a clear separation between possible margination contribution and the suggested "bumping" and "go-with-RBCs" effects.

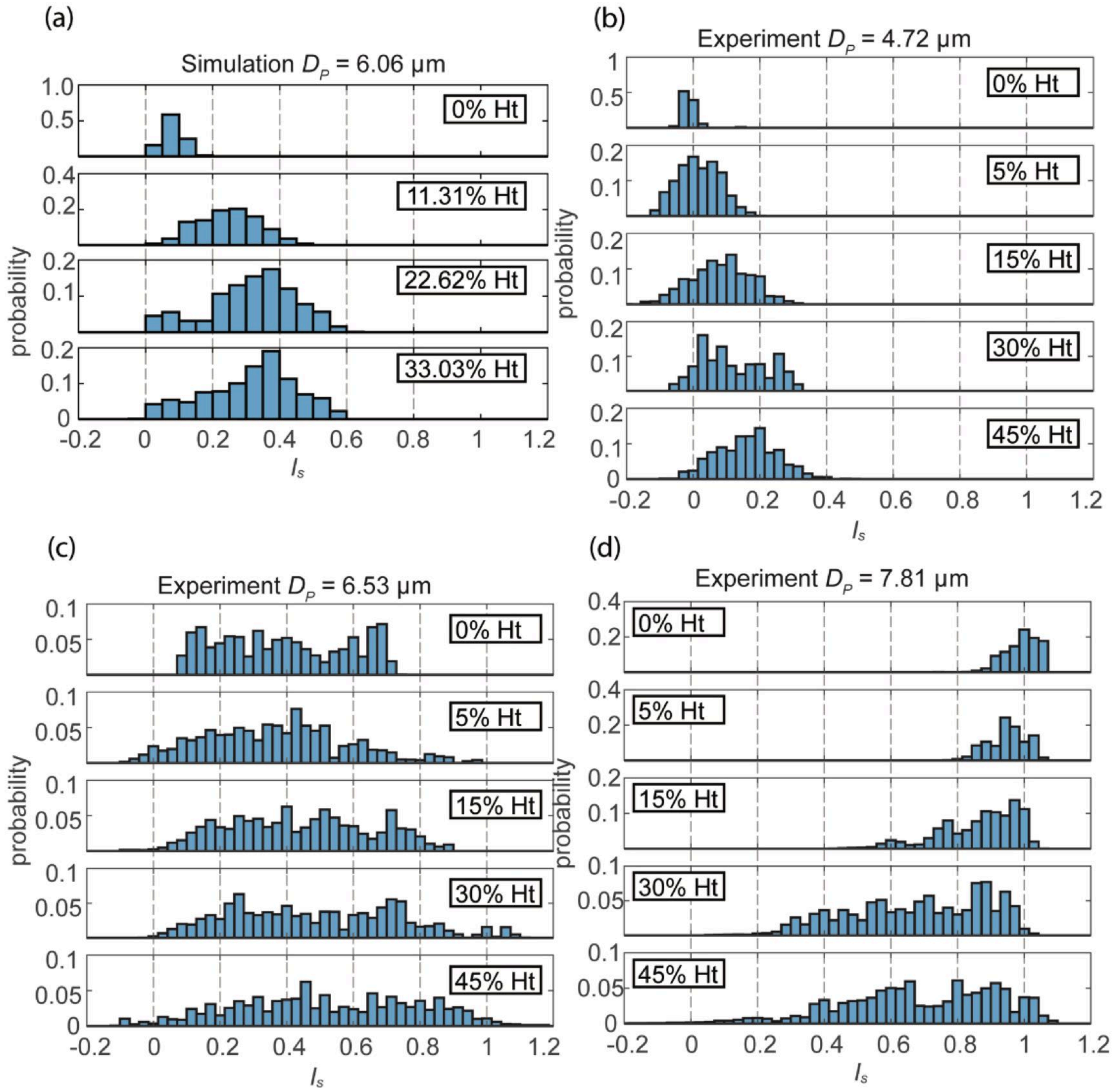

*Figure 10. Outlet distributions from simulations and experiments for different particle diameters as a function of the displacement index $I_S$. Simulation results for $D_P$ = 6.06 μm and experiments for $D_P$ = 4.72 μm compare well qualitatively and both exhibit an increase in the displacement with increasing hematocrit. Experimental data for $D_P$ = 7.81 μm shows a decrease in the particle displacement as hematocrit is elevated. In all cases, outlet distributions generally widen with increasing hematocrit. The outlet distributions for experiments with $D_P$ = 6.53 μm are very broad for all hematocrits (even without RBCs), because the particle diameter is very close to the critical size of the device.*

Another possible contribution, which may affect particle trajectories in the DLD, is altered velocity profile due to an increase of the hematocrit. An altered velocity profile in the DLD can consequently change the critical size of the device. Our simulations show that a change in the average velocity profile is very small for all hematocrits studied, indicating that this effect on the critical size can be practically neglected. This is likely due to the two reasons: (i) we investigate particle motion at moderate hematocrits up to about 33% and (ii) the DLD structure naturally leads to a continuous mixing, preventing the formation of regions with very high hematocrit values. The latter means that the RBC suspension within DLD does not have strong variations in local viscosity and thus, no significant changes in the velocity profile should be expected.

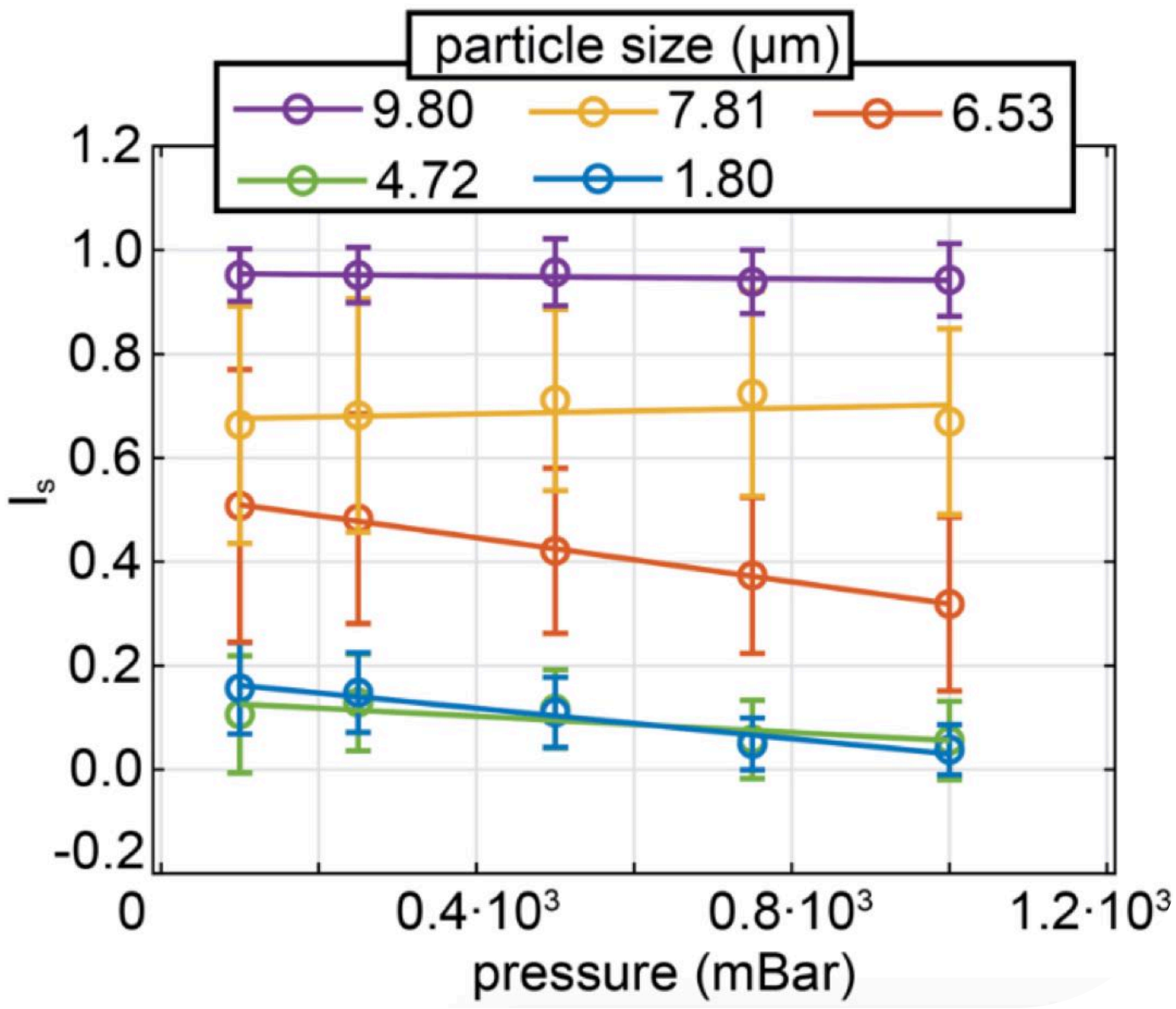

*Figure 11. The effect of flow rate (applied pressure drop) on the displacement index at 45% hematocrit for various particle diameters. Experimental results (device 2, see ESI).*

*Effect of flow rate*

Flow rate through a DLD device is an important parameter since it is directly related to the throughput of the device. Figure 11 shows the effect of flow rate or the applied pressure drop on the displacement index of various particles at 45% hematocrit. For particles with $D_P > D_C$, no significant influence of flow rate on the displacement index is found. This indicates that the 'go-with-RBCs' effect of RBCs, which promotes zigzagging motion, remains unaffected by the flow rate. For particles with $D_P > D_C$, there is a tendency of decreasing displacement index with increasing the pressure drop, indicating that the 'bumping' effect of RBCs, which favors the displacement mode, is reduced. This is likely to occur due to RBC deformation within the DLD device, such that a thinner layer of RBCs near the posts is formed at high flow rates. In conclusion, the effect of flow rate on the displacement index of suspended particles appears to be quite moderate for a wide range of particle sizes, permitting the use of high pressure drops without strong changes in the efficiency of particle sorting or isolation from whole blood. In fact, as the flow rate is increased, some of the reduced separation efficiency due to the high hematocrit is regained, indicating that high throughput separations with good resolution might be easier to achieve by combining high hematocrit with high flow rates.

## Summary and conclusions

Our results show that particles within an RBC suspension or whole blood encounter numerous collisions with the crowding RBCs, which can be roughly divided into two classes on the basis of the corresponding effect on the particle's movement. The first one is a 'bumping' effect due to an increased local hematocrit around the posts (Figure 7), which hinders the particle from entering the lower neighboring streams and favors the displacement motion, as illustrated in Figure 6. The second effect is denoted as 'go-with-RBCs', which forces the particle to flow into the lower neighboring streams and promotes the zigzag motion, as shown in Figure 8. Therefore, the motion of particles in concentrated RBC suspensions is determined by the competition between these two effects and is generally not fully deterministic anymore.

Influence of these two contributions of RBCs on particle trajectories strongly depends on the ratio between particle size and critical size of the device. For large enough particles $D_P >> D_C$, both effects are not able to deteriorate the stable displacement mode of the particles, resulting essentially in no effect of hematocrit on particle trajectory. When the particle size is close to $D_C$ and $D_P > D_C$, the 'go-with-RBCs' effect is able to occasionally force the particles into a zigzag motion such that the particles switch from the pure displacement mode to a mixed mode, as hematocrit is elevated. If the particle size is close to $D_C$ and $D_P < D_C$, the 'bumping' effect of RBCs provides a dominant contribution and the particles exhibit an increase in their displacement index with increasing hematocrit. Finally, for very small particles with $D_P << D_C$, there is nearly no effect of hematocrit with a slight tendency of increasing displacement index at higher hematocrits. Clearly, the closer the diameter of the particles to the critical size of DLD, the larger the effect of RBCs is. In this range of particle diameters, RBCs are able to significantly perturb or disperse the particle trajectory, which would be deterministic without RBCs.

The main result of our study is that the separation of particles at high hematocrits up to whole blood still remains possible, even though RBCs may significantly disperse particle trajectories. The blood primarily influences trajectories of particles, whose sizes are close to the critical size of the DLD array, thereby decreasing the resolution offered by DLD under these conditions. However, certain conditions, when the critical size of a device is far enough from the size of particles of interest, can be advantageous at high hematocrit. For example, for particles with $D_P$ = 7.81μm and $D_P$ = 9.80μm used in our experiments at high hematocrit, the larger particle is nearly unaffected by RBCs and remains in the displacement mode, while the smaller particle exhibits a mixed mode. Therefore, their separate enrichment must be achievable. In conclusion, DLD remains highly promising even at high hematocrits, as we have demonstrated its adept performance for rigid spherical particles. It would also be interesting to test DLD performance for soft particles, since cells such as WBCs in blood are deformable, which will be subject to future research.

## Conflicts of interest

There are no conflicts of interest to declare.

## Acknowledgements

The experimental work and all device processing were carried out within NanoLund at Lund University. We acknowledge funding from FP7-PEOPLE-2013-ITN LAPASO "Label-free particle sorting" (project 607350), NanoLund and the Swedish Research Council (VR) grant no. 2015-05426. Dmitry A. Fedosov acknowledges funding by the Alexander von Humboldt Foundation. We also gratefully acknowledge a CPU time grant by the Jülich Supercomputing Center.